\documentclass[aps,prl,reprint,superscriptaddress,amsmath,amssymb]{revtex4-1}
\usepackage{graphicx}	
\usepackage{bm}	
\usepackage[bookmarks,colorlinks]{hyperref}

\newcommand\figref[1]{\figurename~\ref{#1}}
\begin{document}
\title{Discontinuous Shear Thickening of Frictional Hard-Sphere Suspensions}
\author{Ryohei Seto}
\affiliation{Benjamin Levich Institute, City College of New York, New York, NY 10031, USA}
\author{Romain Mari}
\affiliation{Benjamin Levich Institute,
City College of New York, New York, NY 10031, USA}
\author{Jeffrey F. Morris}
\affiliation{Benjamin Levich Institute, City College of New York, New York, NY 10031, USA}
\affiliation{Department of Chemical Engineering,
City College of New York, New York, NY 10031, USA}
\author{Morton M. Denn}
\affiliation{Benjamin Levich Institute, City College of New York, New York, NY 10031, USA}
\affiliation{Department of Chemical Engineering,
City College of New York, New York, NY 10031, USA}
\date{\today}

\begin{abstract}
Discontinuous shear thickening (DST) observed in many dense athermal
suspensions has proven difficult to understand and 
to reproduce by numerical simulation. 
By introducing a numerical scheme including both relevant hydrodynamic interactions 
and granularlike contacts, 
we show that contact friction is essential for having DST.
Above a critical volume fraction, we observe the existence of two states:
a low viscosity, contactless (hence, frictionless) state,
and a high viscosity frictional shear jammed state.
These two states are separated by a critical shear stress, 
associated with a critical shear rate where DST occurs.
The shear jammed state is reminiscent of the jamming phase of granular matter.
Continuous shear thickening is seen as a lower volume fraction vestige of 
the jamming transition.
\end{abstract}


\maketitle
\begin{figure*}[t]
\includegraphics[width=0.98\textwidth]{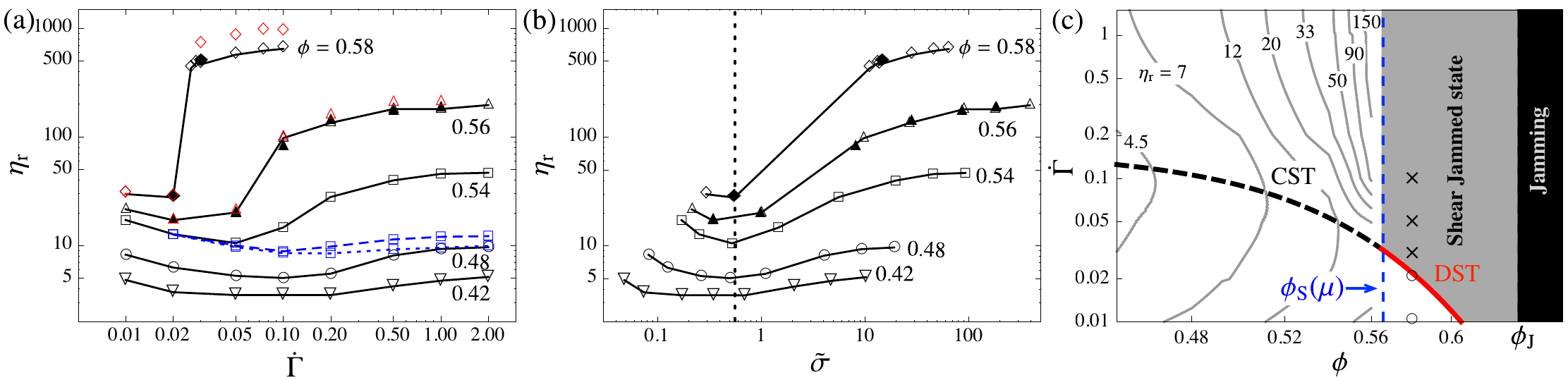}
 \caption{
(color online)
(a), (b) Shear rate and stress dependence 
of the relative viscosity $\eta_{\mathrm{r}}$, respectively.
$\dot{\Gamma}$ is the dimensionless shear rate.
The open and filled symbols indicate the results of $n=512$ and $2048$,
and the volume fractions are shown in the graphs.
The friction coefficient is $\mu = 1$
except for the dashed and dotted blue lines,
for which $\mu =0.1$ and $0$, respectively.
Red symbols show the results with 1.5 times stiffer particles.
(c) DST (red line) and CST (dashed line) are shown in the phase diagram.
The former is expected to reach the jamming point $\phi_{\mathrm{J}}$ for $\dot{\Gamma}\to 0$,
which is not seen because of log scale.
The contour lines for $\phi < 0.56$
are labeled by the relative viscosity, $\eta_{\mathrm{r}}(\phi,\dot{\Gamma})$.
Before jamming (black domain),
the shear jammed states (gray domain) exist.
Observed flowing and jammed states at $\phi=0.58$ 
are shown by circles and crosses, respectively.
}
\label{fig:viscosity}
 \end{figure*}

Suspensions of particles at high volume fraction of solid, 
often termed dense suspensions, have a rich non-Newtonian rheology.
This is particularly striking for the simple system of 
nearly rigid particles in a Newtonian fluid,
which exhibits shear thinning, shear thickening, and normal stresses,
the last associated with strong microstructural distortion,
despite the dominant influence played in such mixtures by viscous 
(Stokes-flow) fluid mechanics~\citep{Morris_2009}.
The phenomenon of discontinuous shear thickening (DST)
(see~\citep{Metzner_1958, Barnes_1989, Mewis_2011, Brown_2012} 
and references therein) is especially fascinating.
Suspensions exhibiting DST flow relatively easily with slow stirring,
but become highly viscous or even seemingly solid 
if one tries to stir them rapidly.
In a rheometer, the transition is seen at a critical shear rate for a given 
volume fraction.
It is often found that DST is completely reversible~%
\citep{Bender_1996}.
DST typically occurs for a volume fraction that exceeds 
a threshold value $\phi_{\mathrm{c}}$, 
which depends on the nature of the suspended particles:
increased nonsphericity or surface roughness seem to lower $\phi_{\mathrm{c}}$.
Continuous shear thickening (CST) is observed below $\phi_{\mathrm{c}}$,
and becomes weaker with decreasing volume fraction.
Although counterintuitive, the abrupt or discontinuous increase 
of viscosity with increase of shear rate is
a generic feature of dense suspensions~\citep{Barnes_1989, Brown_2010},
occurring in both Brownian (colloidal) and non-Brownian suspensions.  
This ubiquity suggests the possibility of a single mechanistic basis 
applicable to the various types of suspension.
DST has yet to be reproduced by a simulation method 
which can unambiguously point to the essential physical features 
necessary for its observation.
This Letter presents a novel method able to identify these features.


Several possible mechanisms have been proposed as the origin of DST.
An order-disorder mechanism~\citep{Hoffman_1972,Hoffman_1974,Hoffman_1998}
describes a low shear rate ordered flow with few interactions between particles
that becomes unstable at high shear rates
and evolves to a disordered, highly interacting viscous flow.
A hydroclustering~\citep{Brady_1985, Brady_1988, Bossis_1989,%
Bender_1995, Bender_1996, Wagner_2009}
or (hydro)contact network~\citep{Melrose_2004a,Melrose_2004} mechanism
attributes the thickening to 
clusters of particles ``glued'' together by the lubrication singularity.
The competition between a force (Brownian or interparticle) tending to
keep particles apart and the imposed shear strain, which tends to push
particles together along the compressional axis, results in narrower
interparticle gaps as the shear rate increases.
The resulting clusters of particles move more rigidly,
effectively increasing the viscosity.
Neither of these scenarios makes a distinction 
between CST and DST, and the development of hydroclusters 
oriented with their dominant principal axis in the compressional quadrant 
in Brownian hard-sphere suspensions leads only to CST~\citep{Foss_2000,Nazockdast_2012a}
even at volume fractions as large as $\phi = 0.58$~\citep{Morris_2002}.
A theoretical approach based on an \textit{ad hoc} mode-coupling theory
attempts to describe DST as a shear-induced glass transition
~\citep{Holmes_2003,Holmes_2005,Cates_2005,Sellitto_2005}.
Another suggested mechanism~%
\citep{Fall_2008, Brown_2009, Brown_2012, Fall_2012}
explicitly relates DST to the existence of an underlying jamming transition
due to the frustration of the granularlike dilatancy by the confining stress.


The appropriate mechanism has been difficult to ascertain.
Most of the mechanisms noted predict the shear rate above 
which shear thickening happens~\citep{Boersma_1990,Maranzano_2001,Brown_2012}.
Experimentally, the order-disorder transition seems 
unnecessary~\citep{Maranzano_2002}, 
at least with a strictly ordered state~\citep{Hoffman_1998}.
Simulations based on purely hydrodynamic modeling, 
such as Stokesian Dynamics~\citep{Brady_1988}, 
show that hydroclusters appear in the semidilute regime
and networks in the concentrated regime ($\phi \gtrsim 0.5$),
where they produce a (weak) CST~%
\citep{Bossis_1989, Phung_1996, Sierou_2002, Melrose_2004a}.
DST has never been reproduced by those models.


A key mechanical issue left largely unconsidered in previous simulation efforts
is the occurrence of contacts, and, in particular, frictional contacts between particles.
%
It is known that, despite the lubrication force, 
particle roughness can lead to contacts,
resulting in qualitative changes from 
the expected behavior of ideal smooth hard particles~%
\citep{Arp_1977, Zhao_2002}.
One consequence of surface contact is an increase of viscosity 
with increasing surface friction~%
\citep{Castle_1996}.
In a colloidal silica suspension exhibiting DST,
increased particle roughness has been shown to lead 
to a smaller critical shear rate~%
\citep{Lootens_2004, Lootens_2005}.
Even for ideally smooth spheres, 
such issues as finite particle deformability may play a role 
for the large stresses that arise at small interparticle gaps.
Such small gaps, dropping to subnanometer scale even for noncolloidal particles, 
lead us to question the relevant physics of close particle interactions.
The experiments cited above, as well as physical intuition,
suggest that contact between particles is an essential ingredient
of the mechanics of flow of highly concentrated suspensions. 

In this Letter we introduce a numerical model 
merging hydrodynamics and features of granular physics.
The model permits contacts between particles 
by assuming a cutoff in the singular resistance 
due to lubrication for a small interparticle gap in the Stokes regime.
These contacts are treated with a model adopted 
from granular physics, involving friction.
Our simulations, limited here to athermal systems
(i.e., not considering Brownian motion, although this may, in principle, be introduced), 
show expected effects of volume fraction and exhibit both CST and DST, with the 
critical ingredient leading to the latter being 
the incorporation of interparticle friction~(\figref{fig:viscosity}).
%


Our model deals with the following interparticle interactions:
the hydrodynamic force $\bm{F}_{\mathrm{H}}$,
the contact force $\bm{F}_{\mathrm{C}}$,
and a repulsive force $\bm{F}_{\mathrm{R}}$.
Since both fluid and particle inertia are neglected, 
the dynamics is overdamped
and forces (and torques) are balanced for each particle:
$\bm{F}_{\mathrm{H}}^{(i)} +\bm{F}_{\mathrm{C}}^{(i)}+\bm{F}_{\mathrm{R}}^{(i)}=0$, 
$i=1, \dotsc, N$.
The hydrodynamic interaction $\bm{F}_{\mathrm{H}}$
in the Stokes regime can be written as a linear function of 
velocities of particles $\bm{U}$ relative to an imposed flow $\bm{U}^{\infty}$
by constructing a resistance matrix $\bm{R}$;
i.e., hydrodynamic forces are of the form 
$\bm{F}_{\mathrm{H}} = -\bm{R} \cdot (\bm{U}- \bm{U}^{\infty})$
(see \citep{Brady_1988,Ball_1997} for details).
The particle velocities can therefore be determined by solving the force balance equations.


For concentrated suspensions,
the resistance matrix can be approximately obtained
by neglecting the far-field or many-body effects
and taking the leading terms of the pair hydrodynamic interactions~\citep{Ball_1997}.
In the simulations, we use the leading terms from the exact solution
for two particles~\citep{Jeffrey_1984,Jeffrey_1992}
in order to handle bidisperse systems,
but the following explanation assumes a monodisperse system for simplicity.
There is a singular factor $1/h^{(i,j)}$ in the resistance to relative motion 
of particles $i$ and $j$, 
where $h^{(i,j)}$ is the interparticle gap.
We argue that it is appropriate, in seeking to represent real suspensions,
to relax the idealization to represent factors 
such as the finite roughness of particle surfaces.
We regularize the lubrication
by inserting a small length $\delta$ to prevent divergence
at contact $h^{(i,j)}=0$ as in \citep{Trulsson_2012}
($\delta =10^{-3}a$ is used, where $a$ is the particle radius).
The squeezing mode of the lubrication force is written as
\begin{equation}
  \bm{F}_{\mathrm{lub}}^{(i,j)}
  = 
  - \alpha(h^{(i,j)})
  (\bm{U}^{(i)} - \bm{U}^{(j)})\cdot \bm{n}^{(i,j)} \bm{n}^{(i,j)}.
  \label{eq:lubforce}
\end{equation}
Here $\alpha(h)= 3 \pi \eta_{\mathrm{0}} a^2/ 2 (h+\delta)$,
where $\eta_0$ is the viscosity of the suspending fluid.
$\bm{n}^{(i,j)}$ is the unit vector along the line of centers 
from particle $i$ to $j$.
Thus, the hydrodynamic force acting on a particle 
is approximately given as
the sum of the regularized lubrication force and the Stokes drag
$\bm{F}_{\mathrm{Stokes}}^{(i)} = 
-6 \pi \eta_0 a \{\bm{U}^{(i)}-\bm{U}^{\infty}(\bm{r}^{(i)})\}$.
The hydrodynamic forces scale with shear rate $\dot{\gamma}$, and
hence there is no essential shear-rate dependence.


The contact force $\bm{F}_{\mathrm{C}}$ is activated for $h^{(i,j)} < 0$.
A simple spring-and-dashpot contact model~\citep{Cundall_1979,Luding_2008,Trulsson_2012} 
is employed to mimic frictional hard spheres;
the normal force is proportional to the overlap $-h^{(i,j)}$:
$\bm{F}_{\mathrm{C,nor}}^{(i,j)} =  k_{\mathrm{n}} h^{(i,j)} \bm{n}^{(i,j)}$,
where $k_{\mathrm{n}}$ is the spring constant.
The friction appears as a tangential force and a torque, 
both proportional to the tangential spring displacement $\bm{\xi}^{(i,j)}$:
$\bm{F}_{\mathrm{C,tan}}^{(i,j)} = k_{\mathrm{t}} \bm{\xi}^{(i,j)}$
and 
$\bm{T}_{\mathrm{C}}^{(i,j)} =  k_{\mathrm{t}} a \bm{n}^{(i,j)} \times \bm{\xi}^{(i,j)}$
(see \cite{Luding_2008} for details),
where $k_{\mathrm{t}}$ is the tangential spring constant.
The tangential force is subject to Coulomb's law
$\bm{F}_{\mathrm{C,tan}} < \mu \bm{F}_{\mathrm{C,nor}}$.
%
%
Even with contact forces, ideal hard-sphere suspensions
should be Newtonian, because different $\dot{\gamma}$ 
result in the same particle trajectories,
but with different time ($\sim 1/\dot{\gamma}$) 
and force ($\sim \dot{\gamma}$) scales .
When trying to mimic hard spheres with linear springs, 
we should avoid introducing an artificial shear-rate dependence.
We therefore choose $k_{\mathrm{n}}$ and $k_{\mathrm{t}} \sim \dot{\gamma}$,
and tune the dashpot resistance 
to keep a short contact relaxation time ($=10^{-3}/\dot{\gamma}$),
in contrast to \cite{Melrose_2004a}.



The shear-rate dependence is introduced by another force
that is not scaled with $\dot{\gamma}$,
which we take as an electrostatic 
double-layer force~\citep{Israelachvili_2011}.
The approximate form 
$\bm{F}_{\mathrm{R}}^{(i,j)} = -C a e^{- \kappa h^{(i,j)} } \bm{n}^{(i,j)} $
is used for $h^{(i,j)}>0$, with $1/\kappa = 0.05a$.
The repulsive force acts to keep particle gaps wider, and is more effective 
at small shear rate, i.e., where the shear time $1/\dot{\gamma}$ is longer.
We thus introduce a dimensionless shear rate
as a ratio of two force scales:
$
\dot{\Gamma}\equiv 6 \pi \eta_0 a^2 \dot{\gamma}/|\bm{F}_{\mathrm{R}}(h=0)|,
$
which is analogous to the P\'eclet number for Brownian suspensions.


Simulations are performed using Lees-Edwards boundary conditions.
The simulation boxes are cubes for $n=512$ particles 
and rectangular parallelepipeds (the shear plane is square, and the depth is one
half of the other dimensions) for $n=2048$.
The influence of particle migration, as previously discussed~\citep{Fall_2010},
can be ruled out here 
since the system is homogeneous owing to the boundary conditions.
A bidisperse system is investigated
to avoid strong ordering  
($a_2 = 1.4 a_1$ and $\phi_1 \approx \phi/2$, i.e., $n_1/n \approx 0.73$),
thereby reducing the potential for an  
order-disorder transition~\citep{Hoffman_1972,Hoffman_1974,Hoffman_1998}.
%


We obtain the dependence of the relative viscosity $\eta_{\mathrm{r}}$
on the dimensionless shear rate $\dot{\Gamma}$ 
and stress $\tilde{\sigma}$ ($\equiv \eta_{\mathrm{r}}\dot{\Gamma}$)
for a range of volume fractions $\phi$ and friction coefficients $\mu$, 
as shown in~\figref{fig:viscosity}.
Two major conclusions can be drawn.
The first is that for frictional spheres, 
a transition from CST to DST is observed upon increase of $\phi$.
For $\mu = 1$, the transition occurs at $0.56 < \phi_{\mathrm{c}} < 0.58$.
Although the shear rate at the onset of thickening 
decreases with increasing $\phi$, 
the onset stress $\sigma_{\mathrm{on}}$ is constant, 
as shown in~\figref{fig:viscosity}\,(b).
This is consistent with experimental observations~%
\citep{Maranzano_2001, Boersma_1990}.
For $\phi > \phi_{\mathrm{c}}$, the high viscosity state 
achieved at high shear rate is strongly dependent on the particle stiffness, 
while no such dependence is seen for $\phi < \phi_{\mathrm{c}}$ 
or in the low viscosity state.
Remarkably, this qualitative difference 
between CST and DST 
has been observed in experiments~\citep{Fall_2008, Fall_2012, Brown_2012}.
There, the control parameter is confinement, 
not particle stiffness, 
but both play the key role for the system to overcome the jammed state:
in experiments by dilation, in simulations by particle overlap.
This has an important conceptual consequence:
if one considers a suspension of 
ideal hard frictional spheres sheared at fixed volume, 
the system becomes solid above the critical shear stress $\sigma_{\mathrm{on}}$, 
as the jammed state achieved is stable against 
any further increment of shear stress.
%

%


The second basic result is the crucial importance of friction.
For $\mu=0.1$, the thickening is substantially weaker than for 
$\mu=1$, and it is completely absent in the frictionless case, 
even at volume fractions approaching the jamming point $\phi_{\mathrm{J}}$.
This means that friction is essential for a shear jammed state to exist.
A similar finding is reported 
for dry granular materials~\citep{Bi_2011, Otsuki_2011}.

\begin{figure*}[tb]
  \includegraphics[width=10cm]{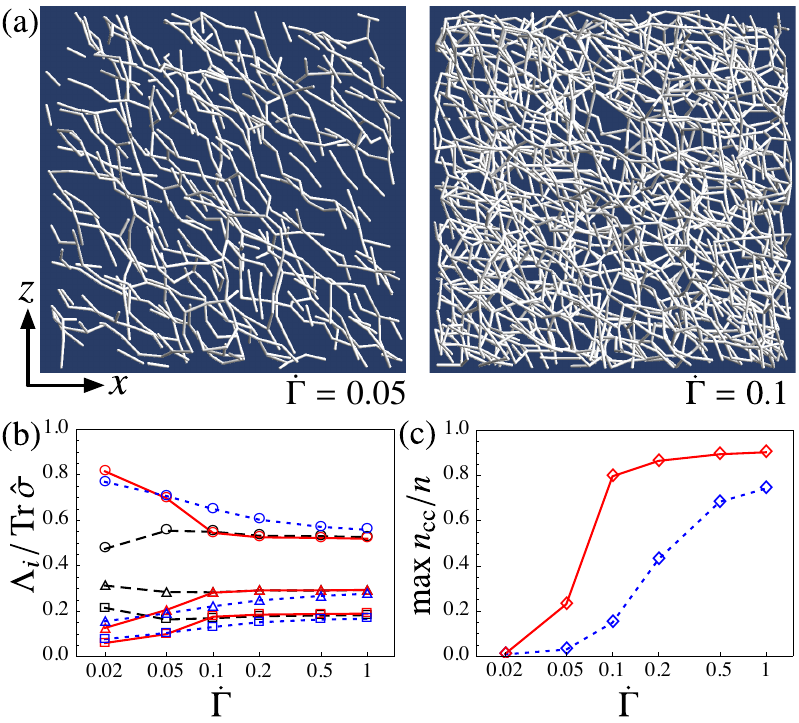}
  \caption{
    (a)
    Particle contacts are visualized as bonds
    at the two shear rates $\dot{\Gamma}=0.05$ and $\dot{\Gamma}=0.1$ 
    exhibiting low and high viscosity states, respectively ($\mu = 1$, $\phi=0.56$, $n=2048$).
    (b)
    The stress eigenvalues,
    $\Lambda_1$(circles), $\Lambda_2$(triangles), and $\Lambda_3$(squares),
    normalized by the trace, are shown ($\phi=0.56$, $n=512$).
    The solid lines (red) indicate the contact stress $\bm{\sigma}^{\mathrm{c}}$,
    and dashed lines (black) the total stress $\bm{\sigma}$ ($\mu = 1$).
    The dotted lines (blue) show the contact stress
    of the frictionless system ($\mu=0$).
    (c)
    The largest contact clusters
    for $\mu=1$ (solid red) and $\mu=0$ (dotted blue) are compared,
    where $n_{\mathrm{cc}}$ is number of particles in a cluster.
  }
  \label{fig:multiaxial_network}
 \end{figure*}

One is then led to think that
shear thickening is related to a shear-induced jamming,
similar to what was suggested in~%
\citep{Cates_1998a,Fall_2008,Brown_2009}.
To test this idea, we show in~\figref{fig:multiaxial_network}\,(a)
the spatial distribution of contact bonds in the system, 
in both the low and high viscosity phases. 
The difference is striking: 
for low shear rate, contact force chains appear,
but only as elongated and isolated objects 
(unique to the combined hydrodynamic-plus-contact force algorithm) along the compression axis,
whereas for high shear rate, a network percolates in all directions.
Similar observations have been made for 
shear jammed states in granular materials~\citep{Bi_2011}.


\citet{Cates_1998a} proposed a simple model of shear jammed states,
describing them by a stress tensor
$ \bm{\sigma} = \Lambda_1 \bm{n}\bm{n} + \Lambda_2 \bm{m}\bm{m} + \Lambda_3 \bm{l}\bm{l}$,
where $\bm{n}$, $\bm{m}$, and $\bm{l}$ 
are the three principal axes of the system, 
corresponding to the three preferential orientations of force chains.
In our simulations, within a one or two degree accuracy,
they coincide with the compression ($\bm{n}$), 
vorticity ($\bm{m}$), and elongation ($\bm{l}$) axes
for all conditions investigated.

We show that the minimal model 
can capture most of the physics of shear thickening,
provided that one distinguishes between the total stress $\bm{\sigma}$ 
and the contribution from frictional contact forces $\bm{\sigma}^{\mathrm{c}}$.
The associated eigenvalues $\Lambda_i$ and $\Lambda_i^{\mathrm{c}}$
normalized by their traces are shown in~\figref{fig:multiaxial_network}\,(b).
It is clear that the result for the total stress shows no difference 
between low and high viscosity states;
the multiaxial stress structure is observed in both states,
meaning that, at such high volume fractions,
the total force chains are percolating in all directions.
However, the shear thickening coincides with a dramatic change in the contact stress.
At low shear rate, 
contacts along the compression axis dominate the stress,
whereas the load is also shared by the other two axes at high shear rate.
The role of friction can be seen by examining the eigenvalues $\Lambda^{\mathrm{c}}_i$ 
in the frictionless contact case, 
as shown in~\figref{fig:multiaxial_network}~(b):
even though the contact stress also gradually evolves 
from a uniaxial to a multiaxial form, no shear thickening is observed.
In addition, the largest contact clusters 
for frictional and frictionless cases are compared
in \figref{fig:multiaxial_network}~(c).
Clearly, friction advances the percolation;
frictional contacts under shear cause local dilatancy,
which compresses remaining gaps.
Thus, friction is essential for the multiaxial contact network to develop sharply, 
i.e., over a narrow range of shear rate, 
and to display the observed mechanical properties.


The percolation of the contact network 
occurs for a minimum shear stress $\sigma_{\mathrm{on}}$,
which is apparently the point 
where repulsive forces among particles are not sufficient 
to prevent the proliferation of contacts.


DST is observed when the percolating network can elastically sustain 
an applied stress.
This is only possible for 
a minimum volume fraction $\phi > \phi_{\mathrm{c}}$,
where there are enough constraints 
to ``lock'' or ``jam'' the structure.
The critical volume fraction
can be identified as the shear jamming~\citep{Bi_2011}:
$\phi_{\mathrm{c}} = \phi_{\mathrm{S}}(\mu)$.
It is close to the 
values observed for the static jamming of frictional spheres, 
but it need not be the same~\citep{Bi_2011}.
When the suspension is forced to flow at high shear rate 
in a strain-controlled experiment, 
the viscosity is dominated by the yield stress 
of the solid network, which is itself influenced by the confinement.
The network of contacts is constantly broken and reformed, 
going from one transient solid configuration to another.


For $ \phi <\phi_{\mathrm{S}}(\mu)$, the CST is a vestige of this jamming transition.
Even when the applied shear stress 
is larger than $\sigma_{\mathrm{on}}$, 
no strictly jammed contact network can form.
Only underconstrained structures appear 
for $\sigma > \sigma_{\mathrm{on}}$ 
(or equivalently $\dot{\Gamma} > \dot{\Gamma}_{\mathrm{c}}$), 
reminiscent of the jammed states seen for $\phi>\phi_{\mathrm{S}}(\mu)$.
These structures still require a large applied stress to flow, 
as they are only deformable via collective rearrangements.
Upon decrease of $\phi$, these networks are increasingly 
underconstrained, and the high viscosity phase fades away.
It is worth noticing that the low viscosity phase, 
essentially frictionless (as there are fewer frictional contacts), 
has similar behavior, 
forming force networks increasingly constrained
as the volume fraction increases~\citep{Lerner_2012,Lerner_2012a}.
It is indeed seen in \figref{fig:viscosity}~(a)
that the viscosity also increases with $\phi$ in this phase. 
However, the point where solid frictionless structures 
appear is only reached for $\phi = \phi_{\mathrm{J}}$,
which is much larger than $\phi_{\mathrm{S}}(\mu)$.
The fact that these two divergences occur at two different volume fractions
is the cause for the blowup of the  difference of viscosity between 
the low shear rate frictionless state and 
the high shear rate frictional state
as $\phi \to \phi_{\mathrm{S}}(\mu)$.


The above results lead us to propose
a schematic phase diagram for the shear thickening 
of athermal suspensions,
represented in the $\phi$-$\dot{\Gamma}$ plane in~\figref{fig:viscosity}~(c).
DST, denoted by a solid (red) line, 
occurs in the range $\phi_{\mathrm{S}}(\mu)<\phi<\phi_{\mathrm{J}}$ for a
critical shear stress $\sigma_{\mathrm{on}}$. 
Asymptotically, in the low viscosity frictionless phase, 
$\eta \propto \left( 1 - \phi/\phi_{\mathrm{J}} \right)^{-q}$ with $ q=2 $~%
\citep{Lerner_2012}, 
which gives for the critical shear rate 
$\dot \Gamma_{\mathrm{c}} \propto \sigma_{\mathrm{on}}
\left(1 - \phi/\phi_{\mathrm{J}}\right)^2$. 
This scaling is, however, difficult to observe in our data range,
as we are still rather far from the divergence.
Above the red line, the shear stress is a yield stress 
of the shear jammed state, proportional to the pressure.
For ideal hard spheres sheared at constant volume, 
this region would simply be inaccessible, 
as the yield stress would be infinite.
Below $\phi_{\mathrm{S}}(\mu)$, CST occurs
around an isostress dashed black line, 
which appears as the continuation 
of the DST red line.
The stress is dominated by the 
proximity of shear jammed states above this dashed line, 
and gives a viscosity 
$\eta \propto \left( 1 - \phi/\phi_{\mathrm{S}}(\mu) \right)^{-q'}$ 
with an estimated $q' \approx 1.5 $.


This phase diagram may well be valid 
even in the case of Brownian suspensions, 
where Brownian motion may play a role similar to the double-layer force, 
namely preventing contacts and reopening gaps at low shear rate.


We thank Ehssan Nazockdast for very useful and stimulating discussions.
J.\,F.\,M.~was supported in part by NSF PREM (DMR 0934206).

%

\end{document}